\documentclass[prl,twocolumn]{revtex4} 
\usepackage{graphicx,amsmath, wasysym}
\pdfoutput=1

\begin{document}

\newcommand{\aver}[1]{\mbox{$<\!#1\!\!>$}}
\newcommand{\bra}[1]{\langle #1|} \newcommand{\ket}[1]{|#1\rangle}
\newcommand{\braket}[2]{\langle #1|#2\rangle}
\renewcommand{\v}[1]{\mathbf{#1}}
\newcommand{\muv}{\mbox{\boldmath$\mu$}}

\def\ua{\uparrow}
\def\da{\downarrow}

\title{Phase-slip oscillator: few-photon non-linearities}

\author{A.~M.~Hriscu}
\affiliation{Kavli Institute of Nanoscience, Delft University of
Technology,\\
PO Box 5046, 2600 GA Delft, The Netherlands}
\author{Yu.~V.~Nazarov}
\affiliation{Kavli Institute of Nanoscience, Delft University of
Technology,\\
PO Box 5046, 2600 GA Delft, The Netherlands}

\date{\today}

\maketitle

{\bf 
Non-linear effects on driven oscillations are important in many fields of physics, ranging from applied mechanics to optics. They are instrumental for quantum applications~\cite{thorne_quantum_1978, slusher_observation_1985}. A limitation is that the non-linearities known up to now are featureless functions of the number of photons $N$ in the oscillator.
Here we show that the non-linearities found in an oscillator where superconducting inductance is subject to coherent phase-slips, are more interesting. They oscillate as a function of number of photons $N$ with a period of the order of $\sqrt{N}$, which is the spread of the coherent state. We prove that such non-linearities result in multiple metastable states encompassing few photons and study oscillatory dependence of various responses of the resonator. A phase-slip process in a superconducting wire is a topological fluctuation of the superconducting order parameter whereby it reaches zero at certain time moment and in certain point of the wire~\cite{zaikin_quantum_1997}.}


Such a process results in a $\pm 2\pi$ change of the superconducting phase difference between the ends of the wire; this produces a voltage pulse. Incoherent thermally-activated phase slips were shown to be responsible for residual resistance of the wire slightly below critical temperature~\cite{skocpol_phase-slip_1974, langer_intrinsic_1967}.
At lower temperatures and in thinner wires phase slips are quantum fluctuations. Although resistance measurements indicate the quantum nature of the phase-slips~\cite{bollinger_2008}, they cannot prove a possible quantum coherence of phase-slip events. A set of other nanodevices~\cite{Mooij2005,mooij_superconducting_2006} have been proposed to verify the coherence experimentally. To facilitate this verification was the initial motivation of our research.  

The inductance $L$ of the wire brings about the inductive energy scale $E_L = \Phi_0/L^2$, where $\Phi_0 =\pi \hbar/ e$ is the flux quantum with $\hbar$ the Planck constant and $e$ the electron charge. It is usually assumed that experimental observation of coherent quantum phase slips requires the phase slip amplitude $E_S$ to be comparable with $E_L$~\cite{mooij_superconducting_2006}. The phase-slip amplitude $E_S$ depends exponentially on the wire parameters, so its value can hardly be predicted and it may be small. This is why it is important to be able to detect arbitrary small values of $E_S$. Our idea is to use a driven oscillator. We prove that in this case the detectable values of $E_S$ are only limited by damping of the oscillator $E_S\approx \hbar \Gamma \ll \hbar \omega_0$. There is an outburst of activity in applying super conducting oscillators for quantum manipulation purposes~\cite{astafiev_single_2007}. The inductance of such an oscillator may be either a thin superconducting wire \cite{wallraff_strong_2004, hofheinz_synthesizing_2009} or a chain of Josephson junctions~\cite{manucharyan_fluxonium:_2009,castellanos-beltran_amplification_2008}. The multi-junction chains also exhibit phase slips and for our purposes are very similar to a wire. Typical experimental values for the main frequency and dissipation rate are $\omega_0 \simeq 10^{10}$ Hz and $\Gamma \simeq 10^5$ Hz.
\begin{figure}[!b] 
\includegraphics[width=0.95\columnwidth]{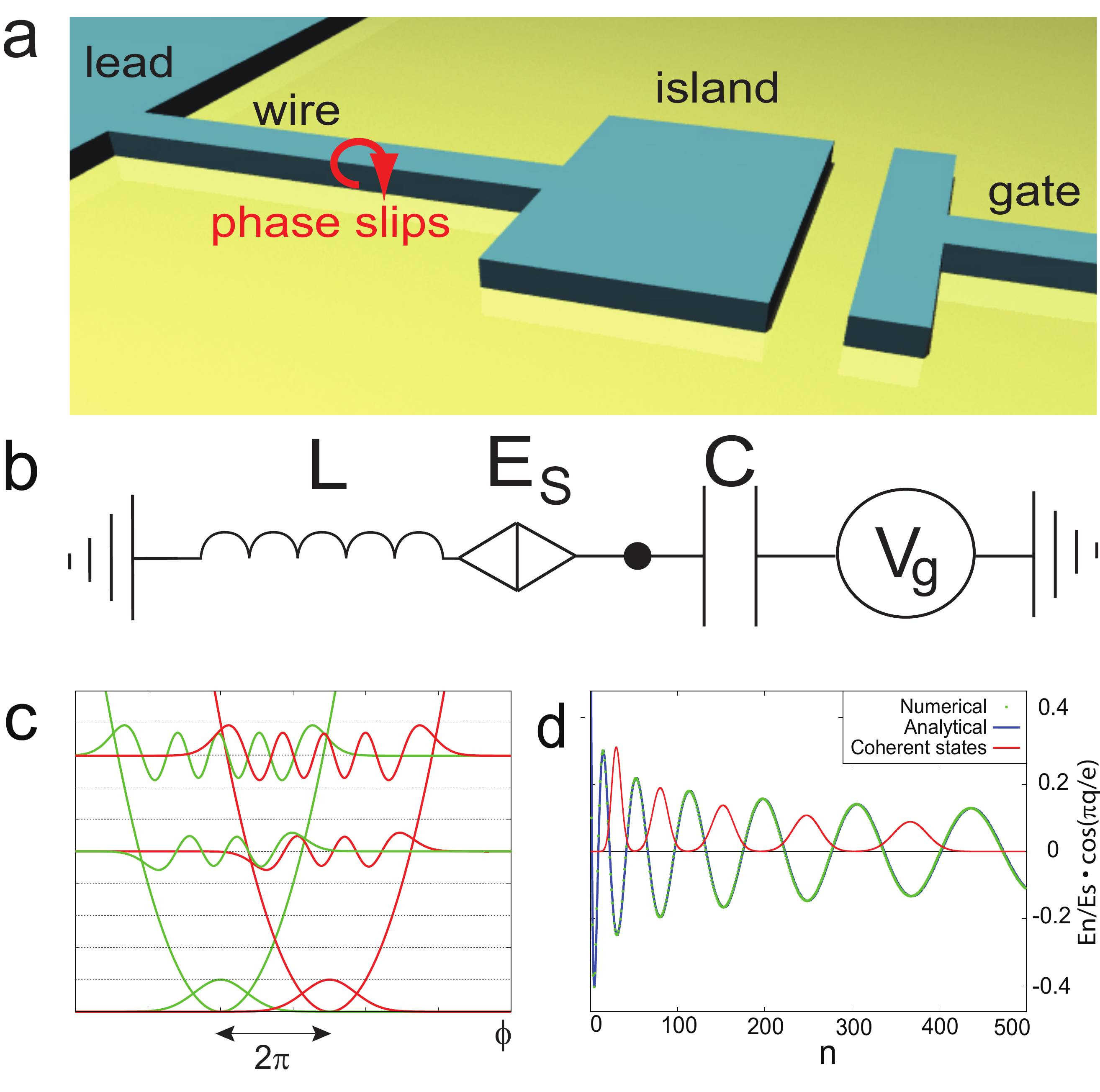}
\caption{\textbf{a}. Phase slip oscillator. A thin superconducting wire connects a lead and an island. The nearby gate electrode induces charge to the island.  The wire is subject to coherent phase-slips. \textbf{b}. The inductance $L$ of the wire and the capacitance $C$ of the island form an oscillator that can be excited with the gate voltage. The crossed diamond represents phase-slips of amplitude $E_S$. \textbf{c}. The phase-slip induced correction to level $n$ is proportional to the overlap integral of the oscillator wave functions (shown here for $n=5;8$). Oscillations of the wave functions give rise to oscillations in $n$. \textbf{d}. The phase-slip corrections to the levels of the oscillator for $\gamma = 0.92$. Exact values (green dots) are fitted with Eq.~\ref{eq:non_linearities}. The photon distribution in several coherent states is plotted in red to illustrate the correspondence between the width of the coherent state and the period of oscillations.}
\label{Fig1}
\end{figure}

This brings us to the system under consideration: the phase slip oscillator. The setup is shown in Fig.~\ref{Fig1}a and the equivalent circuit in Fig.~\ref{Fig1}b. For simplicity, we neglect the effects of the capacitance distribution
along the wire attributing all the capacitance $C$ to the ``island'' at the end of the wire. Without phase slips, the system is a linear $LC$ one-mode oscillator. The a.c.\ component of the gate electrode excites the oscillator while the d.c.\ component induces constant charge $q= C V_g$ to the island. The oscillator is subject to small damping characterised by the energy loss rate $\Gamma \ll \omega_0$, $\omega_0 = 1/\sqrt{LC}$. The dynamics of the oscillator can be described by $\phi$ -- superconducting phase difference dropping along the wire. The phase $\phi$ can take any value and is not restricted to the interval interval $(-\pi, \pi)$. Without phase slips the dynamics are entirely linear with the inductive energy given by $E_L (\phi/2\pi)^2$. So to say, the phase does not know that is supposed to be quantized in units of $2\pi$. Consequently, the charge $q$ on the island does not affect the dynamics. The phase slips shifting the phase by $\pm 2\pi$ can be described by a Hamiltonian acting on the wave function of the system $\Psi(\phi)$\cite{mooij_superconducting_2006}:
\begin{equation}\label{eq:H_S}
\hat{H}_S \Psi(\phi)= \frac{E_S}{2} \sum_{\pm} e^{\pm i \pi q/e} \Psi(\phi \pm 2 \pi).
\end{equation}
The effect of weak phase slips ($E_S \ll \hbar \omega_0$) on the resonantly driven oscillator originates from the shifts $E_n = \bra{n} H_S \ket{n}$ to the otherwise equidistant levels of the oscillator. We immediately see from Eq.~\ref{eq:H_S} that $E_n \propto \cos{(\pi q/e)}$, so the charge induced affects the quantum interference of phase slips with opposite shifts. Any effect of phase slips is thus periodic in gate voltage with a period $2 e/C$. The experimental observation of such dependence unambiguously identifies the quantum coherence of phase slips. As mentioned, we are more interested in the oscillatory dependence on the number of photons $n$. 
One envisages the origin of such a dependence from the fact that the energy shifts $E_n$ are proportional to overlaps of the oscillator wave functions shifted by $\pm 2 \pi$ with respect to each other (see Fig.~\ref{Fig1}c.). The oscillations
of those wave functions which are in phase with a period $\Delta \phi \propto 1/\sqrt{n}$ are thus converted into oscillatory dependence on the photon number. 

\begin{figure}
\centering
\includegraphics[width=\columnwidth]{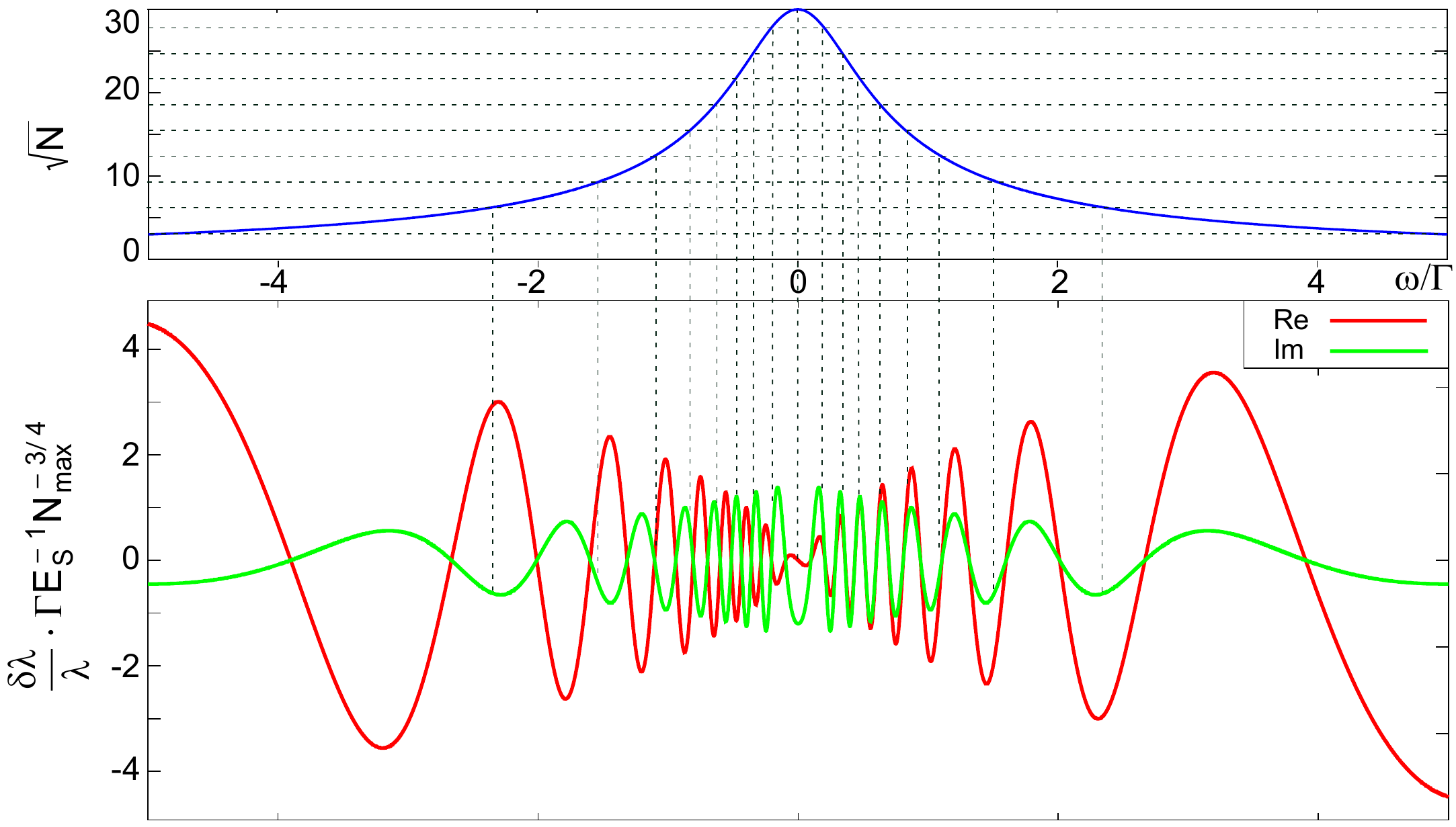}
\caption{Phase slip induced correction $\delta \lambda$ of a driven oscillation versus detuning at $F=15\Gamma$. Real and imaginary parts are shown. Upper pane: $\sqrt{N}$ versus detuning at the same force. It illustrates the oscillation period $\Delta (\sqrt{N})= \pi /\gamma$.}
\label{Fig2}
\end{figure}

The important parameter $\gamma=(2G_QZ/\pi)^{-1/2}$ measures effective impedance of the oscillator $Z = \sqrt{L/C}$, (with $G_Q \equiv e^2/\pi\hbar$) and defines the quantum fluctuations of phase $\propto 1/\gamma^2$. Commonly, electrical resonators have $\gamma \gg 1$. However, superconducting wires provide significant kinetic inductance which may make $\gamma \approx 1$ ~\cite{manucharyan_fluxonium:_2009,castellanos-beltran_amplification_2008}. In this letter, we concentrate on experimentally accessible range $0.3<\gamma<3$. We have found that in this interval the shifts at any $n$ can be sufficiently well approximated by the large $n$ asymptotics:
\begin{equation} \label{eq:non_linearities}
E_n = 2 E_S \cos(\pi q/e) \frac{\cos(2 \gamma \sqrt{n} - \frac{\pi}{4})}{\sqrt{\pi
\gamma} \cdot n^{1/4}} .
\end{equation}
This remarkable is the central point of our findings: it shows that the phase-slips add very unusual non-linearities to the oscillator. The period of oscillations reads $\Delta n = 2\pi \sqrt{n}/\gamma$. At $\gamma \simeq 1$ it compares with the ``width'' $\langle n \rangle =\sqrt{n}$ of a coherent state of the oscillator corresponding to the average number of bosons $n$. For larger $\gamma$, the phase-slip shift $E_n$ will make more oscillations at the scale of the coherent state width. This suppresses the effect of phase slips at large $\gamma$. 

The weak non-linearities induced by phase-slips are neither visible nor important unless the oscillator is resonantly driven. Therefore we include the driving force $2 V_{ac}(t) = \tilde{V} \exp(i (\omega_0 - \omega)t)+ h.c.$, with the detuning  $|\omega|\ll \omega_0$. It is convenient to normalise the driving force such that it enters the Hamiltonian as a combination $\hbar F(\hat{b}+\hat{b}^\dagger)/2$, where $\hat{b}$ and $\hat{b}^\dagger$ are the boson annihilation and creation operators. The force is then $F = e \gamma /\pi \hbar \tilde{V} $. 

In the absence of phase-slips, this force $F$ brings the oscillator 
into a coherent state with the amplitude given by \textit{[reference]}
\begin{equation}
\lambda \equiv \langle b \rangle = -i \frac{F}{\frac{\Gamma}{2} + i\omega}.
\label{eq:zero-order}
\end{equation}
A straightforward but involved perturbation theory (see Supplementary Material) gives the first order correction ($\propto E_S$) to this amplitude, valid at any $\gamma$ and $k_B T$,
\begin{equation}\label{eq:correct_exact}
\delta \lambda = \delta \langle b \rangle = - 2 E_S \cos(\pi q/e)  {\cal E}(\gamma)
\frac{\gamma \lambda^2}{|\lambda| F} J_1(2\gamma |\lambda|),
\end{equation}
with ${\cal E}(\gamma) =\exp ( {-\frac{\gamma ^2}{2 } {\rm cotanh} {\frac{\omega_0}{2 T}}} )$. 
The factor $J_1$ (the first order Bessel function) incorporates oscillations corresponding to the oscillatory behaviour of the energy shifts. The exponential factor ${\cal E}(\gamma)$ is best understood as the effect of averaging of these oscillations over the width of the coherent state. The first order correction is exponentially suppressed at high temperature and $\gamma \gg 1$. While this does not imply that all corrections vanish at $\gamma \gg 1$, we prefer to work at $\gamma \simeq 1$, where the exponent is $\simeq 1$. We will also assume $k_B T \ll \hbar \omega_0$.

In the linear regime, $F \to 0$, the correction amounts to the frequency shift $\hbar\omega \to \hbar \omega - 2 E_S \gamma^2 \cos(\pi q/e) {\cal E}(\gamma)$. The correction becomes noticeable if it is of the order of the line width, $E_{S} \simeq \Gamma$, and can be revealed owing to the oscillatory dependence on gate voltage. 

However, the applicability of the linear regime is limited to almost no photon excited, $\lambda \ll 1$. 
At larger driving, the correction slowly decays with increasing $N \equiv |\lambda|^2$. At $N \gg 1$, the correction becomes significant if $E_S \apprge {\rm max}(\omega,\Gamma) N^{3/4} $. It is interesting to note that the oscillatory correction enhances the dependence on the detuning $\omega$. This is why the correction becomes significant at much smaller $E_S$, $E_S \sim {\rm max}(\omega,\Gamma) N^{1/4}$, if one concentrates on the derivative of the amplitude with respect to detuning, $\partial \lambda / \partial \omega$. We illustrate the scale of the correction and its oscillatory dependence on $\lambda$ in Fig.~\ref{Fig4} and refer to Supplementary Material for details of these estimations.

Let us go beyond perturbation theory, to the regime where the phase-slip correction becomes large, leading to qualitatively different physics. We present a comprehensive semiclassical analysis that captures the essence of the full quantum solution.

\begin{figure}
{\includegraphics[width=\columnwidth]{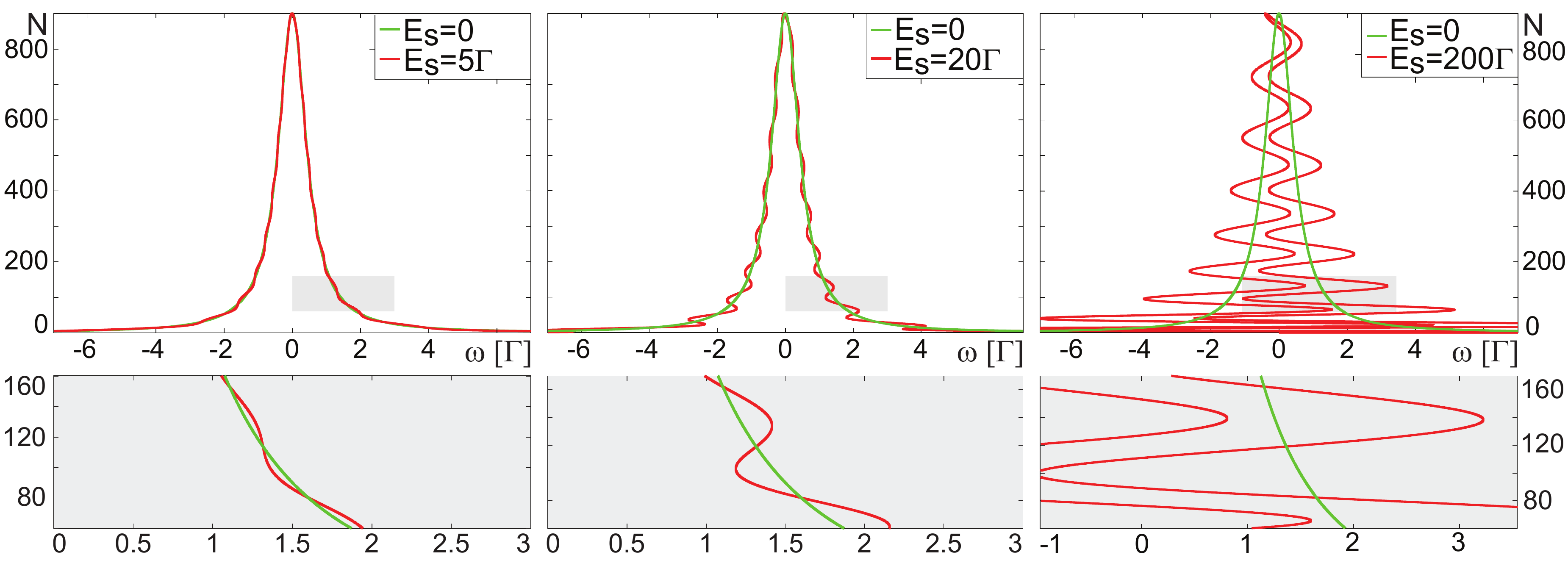} 
\caption{Multiple stability in phase-slip oscillator. We plot the number of photons $N$ versus detuning at $F=15\Gamma$ as predicted by semiclassical approximation (Eq.~\ref{eq:correct_semicls}). In the absence of phase-slips, $N(\omega)$ is a Lorentzian (green line). While at $E_S=5 ~\hbar\Gamma$ the deviation from Lorentzian is barely visible, the enhanced dependence on $dN/d\omega$ brings the oscillator close to the threshold of bistability, as it is shown in inset. At $E_S = 20 ~\hbar \Gamma$ there are already multiple intervals of $\omega$ with two stable configurations. At $E_S = 200 ~\hbar\Gamma$ the curve takes a pronounced corkscrew shape. There is a multitude of stable configurations in all the range of the plot. }
\label{Fig3}}
\end{figure}

\begin{figure}
\includegraphics[width=\columnwidth]{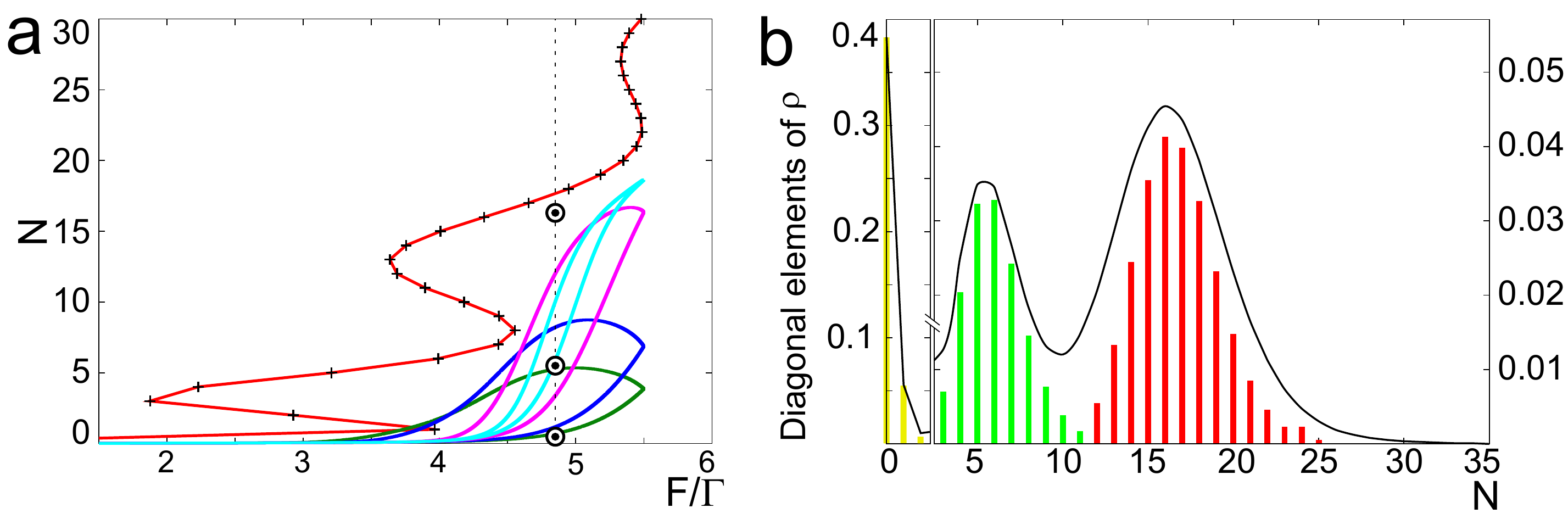} 
\caption{\textbf{a}. Hysteresis in phase-slip oscillator. Crossed curve: semiclassical prediction for $N(F)$ at $\omega=0$. The solid curves give the result of slow sweep of $F$ from $0$ to $5.5\Gamma$. The sweep duration takes values $10^3, 2\cdot 10^3,10^4$ and $4\cdot 10^4~ \Gamma^{-1}$ from lowest to uppermost curve. The pronounced hysteresis indicates exponentially long life-times of the metastable states. Black circles indicate the pure states that contribute to the equilibrium density matrix at $F=4.85 \Gamma$ (in detail in Fig.b). \textbf{b}. Pure states. We show diagonal elements of the equilibrium density matrix at $F=4.85 \Gamma$. The pulses of different colour show contributions to diagonal elements from three pure states of highest probability. We see a ``dark state'' with predominantly zero number of photons and two coherent-like states centred around 5.5 and 16.3 photons.}
\label{Fig4}
\end{figure}

In the semiclassical approximation, we replace $N$ by a continuous variable. The non-linearities modify the detuning $\omega$ in Eq.~\ref{eq:zero-order}, $\omega \to \omega + dE(N)/\hbar  dN$, where $E(N)$ is defined by Eq.~\ref{eq:non_linearities} at $\gamma \approx 1$. Squaring Eq.~\ref{eq:zero-order} yields a self-consistency equation for $N$ at given $F$ and $\omega$~\cite{Cohen-Tannoudji1992}: 
\begin{equation}\label{eq:correct_semicls}
N =  \frac{F^2}{ (\frac{\Gamma}{2})^2 + (\hbar \frac{dE}{dN} + \omega )^2 }.
\end{equation} 
That suffices to make implicit plots $N(F, \omega)$. For common non-linearities $dE/dN$ is $\propto N$. This gives either a single solution for $N(\omega)$ or three solutions corresponding to two metastable states. The oscillatory dependence on $N$ changes this drastically.  To elucidate, we plot $N(\omega)$ in Fig.~\ref{Fig3} at fixed $F=15 \Gamma$. At negligible $E_S$, $N(\omega)$ is a Lorentzian. Phase-slip corrections shift the curve horizontally, the magnitude of the shift oscillating with a period $\simeq \sqrt{N}$. At sufficiently large $E_S$ this results in impressive characteristic ``corkscrew'' shape. At any given $\omega$ within the oscillator line width one finds a multitude of states that differ in $N$. About half of these states are stable. We stress the tunability of this phase-slip oscillator: small changes of the driving force, detuning, or charge induced change the number of stable states, thereby enabling easy manipulation of $N$. 

Generally, one expects fluctuation-induced switching between the available stable states. The semiclassical analysis does not account for that. Nor does it prove if a given metastable solution corresponds to a pure quantum state. It is also not clear if the semiclassical prediction for the metastable solutions works for the states with few photons. To understand this, we have performed numerical simulations with the full quantum equation Eq.~\ref{eq:quantum} for density matrix. 
For illustration, we set $\omega=0$ and $E_S$ to a moderate value of $6 ~\hbar\Gamma$. We initialise the density matrix to vacuum, $\ket{0} \bra{0}$, and compute its time-dependence while making a linear sweep of $F$ from $0$ to $5.5~ \Gamma$ and back. From semiclassics we expect up to 3 metastable states in this force interval. Plotting $\langle n \rangle$ versus $F$ for different sweep durations $T$ gives a series of curves with evident hysteresis (see Fig.~\ref{Fig4}a). Generally, one expects the relaxation time of the density matrix to be of the order of $1/ \Gamma$. Remarkably, a noticeable hysteresis persists even at time intervals $4 \cdot 10^4 ~\Gamma^{-1}$. This clearly indicates an exponentially long life-time of the metastable states even for few photons. We hypothesise that the oscillator spends most of the time in one of such states while  rare switching between these states result in equilibration of the probabilities to be in these states. Such equilibration occurs at the time scale corresponding to the slowest switching rate.
To prove this illustratively, we have computed the equilibrium density matrix at $F= 4.85~\Gamma$ and expanded it into a sum over orthogonal quantum states. We have found that the density matrix is mainly contributed by three pure states: one ``dark'' state $\approx \ket{0}$ and two coherent-like state centred around $5.5$ and $16.3$ photons respectively with probabilities $0.46, 0.25$ and $0.15$. The remaining probability corresponds to ``excited'' states that have nodes at positions of the coherent-like states centred at $5.6, 16.3$. The relaxation time that characterises the slow switching is $300~\Gamma^{-1}$ at this value of $F$. About $4000$ photons are absorbed and emitted in the oscillator during this time interval; this proves the extraordinary robustness of the states involved. 

To conclude, we have investigated the effect of non-linearities produced in a superconducting resonator by coherent phase-slips. These non-linearities are very distinct from those previously known owing to the oscillatory dependence on number of photons with a period $\simeq \sqrt{N}$. We have demonstrated that at semiclassical level the non-linearities result in a multitude of metastable states. The position and number of these metastable states can easily be tuned by changing the driving force. At quantum level, we have demonstrated that there is a single quantum state corresponding to the semiclassical metastable states. These states are robust with an exponentially long switching time although they encompass only few photons. These features of the \textit{phase-slip oscillator} may be applied for a wide range of applications, as ultra-sensitive measurements, quantum manipulation and an unambiguous experimental verification of coherent phase-slips. 

We are grateful to J.E. Mooij, members of his team and to A. Ustinov for useful discussions. This work is supported by the `Stichting voor Fundamenteel Onderzoek der Materie (FOM)' and the `Nederlandse Organisatie voor Wetenschappelijk Onderzoek (NWO)'.

\section*{Methods}
We start with the Hamiltonian of the driven oscillator,
\begin{equation*}
\hat {H} = \hbar \omega_0 b^{\dagger}b + {\rm Re} \{ F e^{i(\omega_0 - \omega)t}\} \frac{b^{\dagger}+b}{2} + \hat{H}_S,
\end{equation*}
where the phase-slip term $H_S$ is given by Eq.~\ref{eq:H_S}.

We implement the rotating-wave approximation to arrive to the equation for density matrix $\hat{\rho}$ :
\begin{equation}\label{eq:quantum}
\frac{\partial \hat{\rho}}{\partial t} = - \frac{i}{\hbar} [\hat{H}_R, \hat{\rho}] + 
 \Gamma \Big( b \hat{\rho} b ^{\dagger} - \frac{1}{2} (b^{\dagger}b\hat{\rho} + \hat{\rho} b^{\dagger}b) \Big),
\end{equation}
where the terms including the dissipation $\Gamma$ in Eq.~\ref{eq:quantum} are of conventional form~\cite{Cohen-Tannoudji1992} (assuming $k_BT \ll \hbar\omega_0$) and 
\begin{equation*}
\hat{H}_R = E(b^{\dagger} b) + \frac{F b^{\dagger}+ F^{*} b}{2} + \omega b^{\dagger} b .
\end{equation*}
Here $E(n) = \bra{n} \hat{H} \ket{n}$ are the energy shifts induced by the phase-slips. These shifts are expressed through the hypergeometric function $_1F_1$:
\begin{equation*}
E_n = 2 E_S \cos(\pi q/e) \cdot \exp{(- \gamma^2/2)} _1{\rm F}_1[-n,1, \gamma^2].
\end{equation*}
Large-$n$ asymptotics are given by Eq.~\ref{eq:non_linearities}. 
Without phase-slips ($E_n = 0$), the solution of Eq. (\ref{eq:quantum}) for long time intervals reads
$$
\hat{\rho}(t \rightarrow \infty) = \exp {\Big[ - \frac{\hbar \omega_0}{k_B T} \big (b^{\dagger}b- \lambda b^{\dagger} + \lambda^{*} b +|\lambda|^2 \big) \Big] } ,
$$
where $\lambda \equiv \langle b \rangle$ is defined by Eq.~\ref{eq:zero-order}. To compute the first order correction $\delta \lambda$, we develop a rather involved perturbation theory for the density matrix that is outlined in the Supplementary Material, which leads to Eq.~\ref{eq:correct_exact}.

For our numerical calculations, we solve the evolution equation Eq.~\ref{eq:quantum} by $4$th order Runge-Kutta method with a fine time step($\approx 0.02 ~1/\Gamma $). Typically, we take into account $N = 50$ states. Since we investigate slow dynamics induced by rare switching, the simulation runs take rather long time ($\approx 10$h for $T \approx 4\cdot 10^3~1/\Gamma$).


\onecolumngrid


\begin{center}
\textbf{{\large Supplementary Material for\\
``Phase-slip oscillator: few-photon non-linearities''}}\\
\bigskip
A.~M.~\ Hriscu, Yu.~V.~\ Nazarov\\
\textit{Kavli Institute of NanoScience, Delft University of Technology, 2628 CJ Delft, The Netherlands}
\end{center}

\section{Perturbation theory for density matrix}
Here we present a more detailed account of how we have calculated the first order correction to the oscillator amplitude (Eq.~ \ref{eq:correct_exact}) using perturbation theory for the density matrix that obeys Eq.~\ref{eq:quantum}. We rewrite it in compact form using a super-operator $\hat{\hat{\Sigma}}$ that accounts for dissipation, detuning and the driving force:
$$
\frac{\partial \rho}{\partial t} = \hat{\hat{\Sigma}} \rho - \frac{i}{\hbar} [H_S, \rho]
$$
The stationary solution in the absence of phase-slips ($E_S = 0$) reads
$$
\hat{\rho}_{\infty} = \exp {\Big[ - \frac{\hbar \omega_0}{k_B T} \big (b^{\dagger}b- \lambda b^{\dagger} + \lambda^{\star} b +|\lambda|^2 \big) \Big] },
$$
where $\lambda \equiv \langle b \rangle$ is defined by Eq.~\ref{eq:zero-order}. The phase-slip Hamiltonian $E(b^{\dagger}b)$, Eq.~\ref{eq:H_S}, is treated as a perturbation. The first order correction to the density matrix satisfies the following condition
\begin{equation}\label{eq:supl_rho}
\frac{\partial \rho^{(1)}}{\partial t} = \hat{\hat{\Sigma}} \rho^{(1)} - \frac{i}{\hbar} [E(b^{\dagger}b), \rho_{\infty}].
\end{equation}
This equation can be formally solved by making use of the superoperator resolvent $\hat{\hat{R}} = (\partial/\partial t - \hat{\hat{\Sigma}})^{-1}$ :
$$
\rho^{(1)} (t) = - \frac{i}{\hbar}\int_{- \infty}^{t} dt' \hat{\hat{R}}(t-t') [E(b^{\dagger}b), \rho_{\infty}].
$$
To avoid explicit evaluation of this cumbersome resolvent we re-write it as an integral over a time-dependent operator $\tilde{\rho}(\tau)$,
$$
\rho^{(1)} (t) = \int_{0}^{\infty} d\tau \tilde{\rho} (\tau) , \quad \tilde{\rho}(\tau) = -\frac{i}{\hbar}\hat{\hat{R}}(\tau) [E(b^{\dagger}b), \rho_{\infty}].
$$
The thus defined $\tilde{\rho}(\tau)$ obeys the time evolution equation 
\begin{equation}\label{eq:supl_rhotilde}
\frac{\partial \tilde{\rho}(\tau)}{d \tau} = -\frac{i}{\hbar} [\hat{\hat{\Sigma}}, \tilde{\rho}(\tau)],
\end{equation}
with the initial condition
$$
\tilde{\rho}(\tau=0) = -\frac{i}{\hbar} [E(b^{\dagger} b), \rho_{\infty}].
$$
Let us now represent the quantity $E(b^{\dagger}b)$ via an integral that singles out the diagonal elements of the original phase-slip operator:
$$
E(b^{\dagger}b) = \int_{0}^{2 \pi} \frac{d \chi}{2 \pi} e^{i \chi b^{\dagger}b} H_S e^{-{i \chi b^{\dagger}b}}.
$$
This operator consists of two shift operators that are expressed in terms of boson operators as $ e^{\pm i \gamma(b^{\dagger} + b)}$. Therefore 
$$
E(b^{\dagger}b) = E_S \int_{0}^{2 \pi} \frac{d \chi}{2 \pi} \Big[e^{i \pi q/e} e^{i \gamma(b^{\dagger}e^{i\chi } + be^{-i\chi })} + e^{-i \pi q/e} e^{-i \gamma(b^{\dagger}e^{i\chi } + be^{-i\chi })} \Big].
$$
Let us concentrate on the first order correction to the amplitude due to the $2 \pi$ phase-slip (corresponding to terms $\propto e^{i \pi q/e}$). It is expressed as
$$
\delta \langle b \rangle = \textrm{Tr}(b \rho^{(1)}) =  \int_{0}^{ \infty} d\tau \textrm{Tr} [b \tilde{\rho}(\tau)] =- \frac{i}{\hbar} E_S  e^{i \pi q/e} \int_{0} ^{\infty} d\tau \int_0 ^{2 \pi} d \chi \textrm{Tr} (b ( \tilde{\rho}_{+}(\tau, \chi) - \tilde{\rho}_{-}(\tau, \chi) )) .
$$
Here, $\tilde{\rho}_{\pm}(\tau, \chi)$ satisfies Eq.~\ref{eq:supl_rhotilde} with initial conditions $ \tilde{\rho}_{+}(0, \chi) = e^{ i \gamma(b^{\dagger}e^{i\chi } + be^{-i\chi })} \rho_{\infty}$ and $ \tilde{\rho}_{-}(0, \chi) = \rho_{\infty} e^{i \gamma(b^{\dagger}e^{i\chi } + be^{-i\chi })} $. 

We concentrate on the quantities $\xi_{\pm}$ defined as
$$
\xi_{\pm} (\tau, \chi) =  \frac{\textrm{Tr} (b \tilde{\rho}_{\pm}(\tau, \chi))}{T(\chi,\gamma)} , \quad T(\chi, \gamma)=\textrm{Tr}(\tilde{\rho}_{\pm}(0, \chi)).
$$
They obey the equation $\partial \xi_{\pm}/ \partial \tau = - i \omega \xi_{\pm} -i F - \xi_{\pm} \Gamma/2$, the same equation as the time dependent amplitude of the oscillations obeys. This yields
$$
\xi_{\pm} (\tau, \chi) = \xi^{\infty}(\gamma) + (\xi_{\pm}^0(\chi; \gamma) - \xi^{\infty}) e^{-(\frac{\Gamma}{2}+ i \omega)\tau},
$$
with 
$$ 
\xi^{\infty}(\gamma) = -i \frac{F}{\frac{\Gamma}{2}+ i \omega} \quad \textrm{and} \quad \xi_{\pm} (0, \chi) =  \frac{\textrm{Tr} (b \tilde{\rho}_{\pm}(0, \chi))}{T(\chi,\gamma)}.
$$
The integral over the time variable $\tau$ gives a factor of $\frac{\Gamma}{2}+ i \omega$ in the denominator. Let us also note that 
$$
 \xi^0_+ - \xi^0_- = \frac{\textrm{Tr}([b, e^{i \gamma(b^{\dagger} e^{i \chi}+ b e^{-i \chi})}]\cdot \rho_{\infty})}{T(\chi,\gamma)} = i \gamma e^{i \chi}.
$$
Combining this with the contribution of the $-2 \pi$ phase-slips, we obtain
$$
\delta \langle b \rangle = \frac{E_S/\hbar}{\frac{\Gamma}{2} + i\omega} \gamma\int_{0}^{2 \pi} \frac{d \chi}{2 \pi} e^{i \chi} [e^{i \pi q/e} T(\chi,\gamma) - e^{-i \pi q/e} T(\chi, -\gamma)].
$$
One computes the traces involved by making use of the expression for $\rho_{\infty}$.
$$
e^{i \pi q/e} T(\chi,\gamma) - e^{-i \pi q/e} T(\chi, -\gamma) =  \cos(\pi q/e) \exp\left[-\frac{\gamma^2}{2} {\rm cth}\left(\frac{\alpha}{2}\right)
+i \gamma {\rm Re} \left( e^{i\chi} \lambda^{*}\right) \right],
$$
where $\alpha = \frac{\hbar \omega_0}{k_B T} $.
To  complete the calculation, we shift $\chi$ with the phase $\zeta$ of $\lambda^{*}$ to obtain
$$
\langle b \rangle = \frac{E_S/\hbar  \cos(\pi q/e)}{\frac{\Gamma}{2} + i\omega} \exp\left[-\frac{\gamma^2}{2} {\rm cth}\left(\frac{\alpha}{2}\right)\right]\frac{|\lambda|}{\lambda^{*}} \gamma\int \frac{d \chi}{2\pi} 2 \cos(\chi) \exp\left[2 i \gamma|\lambda|\cos(\chi)\right].
$$
Integration over $\chi$ yields the final result,
$$
\langle b \rangle = \frac{2iE_S \cos(\pi q/e) /\hbar }{\frac{\Gamma}{2} + i\omega} \exp\left(-\frac{\gamma^2}{2} {\rm cth}\left(\frac{\alpha}{2}\right)\right)\frac{|\lambda|}{\lambda^{*}} \gamma
J_1(2\gamma|\lambda|).
$$
This is the oscillating function of $q$ and $\lambda$ Eq.~\ref{eq:correct_exact} discussed in the main text.

\section{ESTIMATIONS OF THE RELATIVE MAGNITUDE OF THE CORRECTION}
In this section we show how we have obtained the estimates for the phase-slip amplitude that causes the significant correction to the amplitude of driven oscillations. We use Eq.~\ref{eq:correct_exact} and consider several limits. In the first limit, $\gamma |\lambda| \ll 1 $, the first order Bessel function is expanded as $J_1(x) \approx \frac{x}{2}$. Then the correction becomes significant if
$$
E_S \gtrsim \frac{\textrm{max}(\omega, \Gamma/2)}{\gamma^2}.
$$
In the opposite limit, $\gamma |\lambda| \gg 1 $, the Bessel function is approximated
by the cosine: $J_1(x)\approx \sqrt{\frac{2}{\pi x}} \cos(x)$, and therefore the correction becomes significant if
$$
E_S \gtrsim \frac{\textrm{max}(\omega, \Gamma/2)}{\gamma^{1/2}} N^{3/4},
$$
where $N$ is the average number of excited photons. The correction relatively decreases with increasing $N$ and in the limit $N\gg 1$ it oscillates.

As mentioned in the main text, the oscillations enhance the frequency dependence of the correction. This feature might be used for experimental detection of the correction. To prove this, let us consider the derivative of the correction with respect to $\omega$ and compare it with the same derivative of the zero-order amplitude:
$$
\frac{d \delta \lambda}{d \omega} / \frac{d |\lambda|}{d \omega}
\propto \frac{E_S}{F} \sqrt{|\lambda|} \gamma^{3/2} \sin(2 \gamma
|\lambda|) 
$$
By analysing this we understand that the correction becomes significant if 
$$
E_S \gtrsim \frac{\textrm{max}(\omega,\Gamma/2)}{\gamma^{3/2}} N^{1/4}
$$
We see that this improves the estimation of the significant phase-slip amplitude by a factor of $\sqrt{N}$.

To summarise, we have three estimations for the significant phase-slip amplitudes. In increasing order, at $E_S \simeq \Gamma$, the correction is significant in the linear response regime $N \ll 1$. At $E_S \sim \Gamma N^{1/4}$ one sees the significant effect on the frequency dependence of the oscillation amplitude $\langle b\rangle$. At $E_S \sim \Gamma N^{3/4}$, the overall change of the magnitude becomes dominant.

\begin{figure}
\includegraphics[width=0.45\columnwidth]{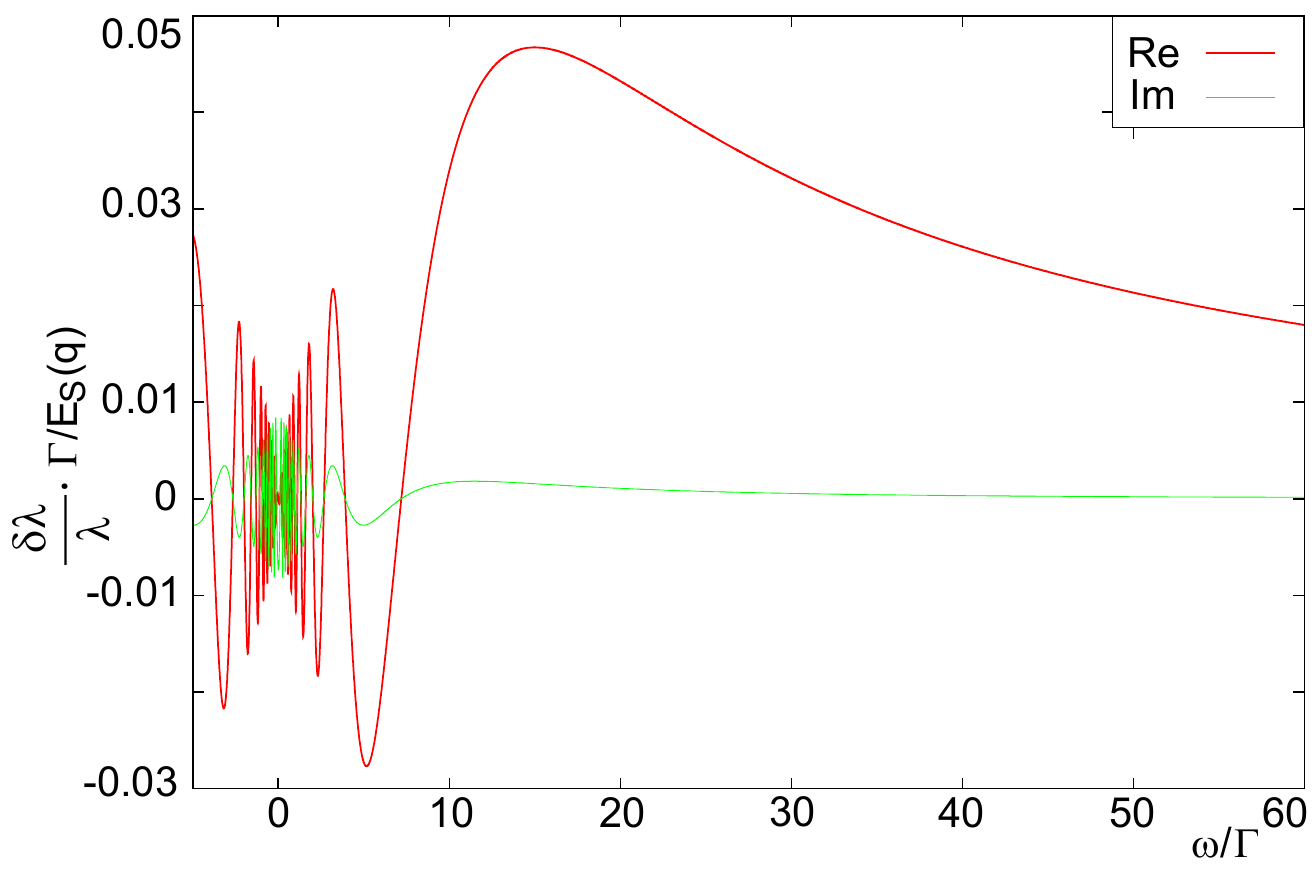}
\caption{The real and imaginary parts of the relative correction $\delta \lambda/\lambda$. $E_S(q) = E_S \cos(\pi q/e)$. A zoom of this plot for the range $\Gamma/\omega =[ -5, 5]$ has been shown in the main text in Fig.~\ref{Fig2}. The real part of the correction reaches maximum at $\omega \approx F$.}
\label{fig:correction_plt} 
\end{figure}

\begin{figure}
\centering
\includegraphics[width=0.90\columnwidth]{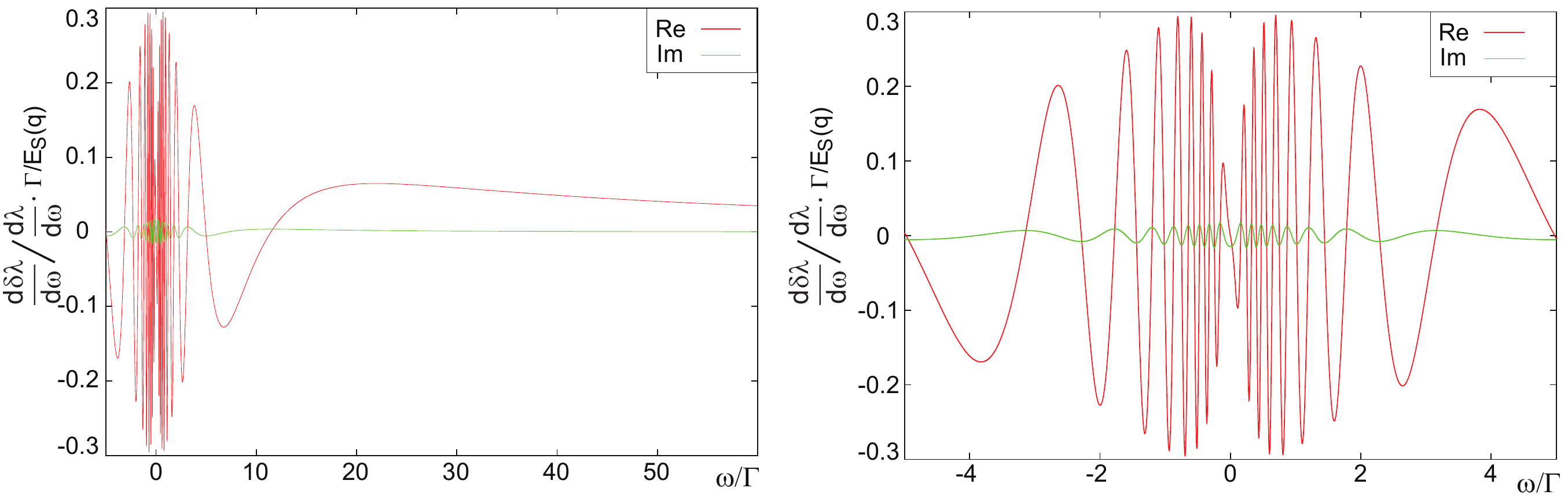}
\caption{Left panel: the derivative of the correction with respect to the frequency, $\frac{d (\delta \lambda )}{d \omega}/
\frac{d \lambda}{d \omega}$, in units of $E_S(q)/\Gamma$, with $E_S(q) = E_S \cos(\pi q/e)$ as function of $\omega/\Gamma$, at fixed driving force of $F=15~\hbar \Gamma$. Right panel: zoom of Fig.~\ref{fig:derivative} for small detunings.}
\label{fig:derivative}
\end{figure}

We illustrate these conclusions with plots. We plot the relative magnitude of the correction versus detuning in all the figures, at a fixed value of the a.c.~ driving force $F=15{\Gamma}$. This corresponds to $N=900$ excited photons at zero detuning. At larger detunings this number decreases reaching $N \simeq 1$ at $\omega \simeq F$. 
Fig.~\ref{fig:correction_plt} represents the real and imaginary part of the correction $\delta \langle b \rangle /\langle b \rangle$ from Eq.~\ref{eq:correct_exact}. The real part of the correction reaches its maximum at $\omega \approx F$, as seen in Fig.~\ref{fig:correction_plt}, while the number of photons is small $N \simeq 1$. Therefore one does not need to excite the oscillator in higher states to get large enough signal. At small frequencies it oscillates strongly with smaller amplitude, while at larger frequencies it decays to zero as $1/{\omega}$. The imaginary part oscillates also strongly at small frequencies, but it goes to zero much faster, $1/{\omega}^2$. We also notice that the real part is odd in frequency, while the imaginary part is even. Their oscillations are in phase for $\omega >0$ and opposite in phase at $\omega <0$.
Fig.~\ref{fig:derivative} shows the plot of the ratio of derivatives with respect to the frequency. We see how this enhances the relative magnitude of the corrections at small detunings (large $N$).

\hfill

\end{document}